\documentclass{article}
\usepackage{graphicx}
\usepackage{amsmath}
\usepackage{amssymb}
\begin{document}
\renewcommand\Re{\operatorname{Re}}
\renewcommand\Im{\operatorname{Im}}
 
\begin{center}
    {\Huge{A Comparison of Frequency Downshift Models of Wave Trains on Deep Water}}\\[10pt]
    {\Large{John D.~Carter}}\\[4pt]
    {\Large{Diane Henderson}}\\[4pt]
    {\Large{Isabelle Butterfield}}\\[10pt]
    {\large{\today}}
    \end{center}

\date{\today}

\section*{Abstract}
Frequency downshift (FD) in wave trains on deep water occurs when a measure of the frequency, typically the spectral peak or the spectral mean, decreases as the waves travel down a tank or across the ocean.  Many FD models rely on wind or wave breaking.  We consider seven models that do not include these effects and compare their predictions with four sets of experiments that also do not include these effects.  The models are the
(i) nonlinear Schr\"odinger equation (NLS), 
(ii) dissipative NLS equation (dNLS), 
(iii) Dysthe equation, 
(iv) viscous Dysthe equation (vDysthe), 
(v) Gordon equation (Gordon) (which has a free parameter), 
(vi) Islas-Schober equation (IS) (which has a free parameter), and 
(vii) a new model, the dissipative Gramstad-Trulsen (dGT) equation.  The dGT equation has no free parameters and addresses some of the difficulties associated with the Dysthe and vDysthe equations.  We compare a measure of overall error and the evolution of the spectral amplitudes, mean, and peak.  We find: 
(i) The NLS and Dysthe equations do not accurately predict the measured spectral amplitudes. 
(ii) The Gordon equation, which is a successful model of FD in optics, does not accurately model FD in water waves, regardless of the choice of free parameter. 
(iii) The dNLS, vDysthe, dGT, and IS (with optimized free parameter) models all do a reasonable job predicting the measured spectral amplitudes, but none captures all spectral evolutions. 
(iv) The vDysthe, dGT, and IS (with optimized free parameter) models do the best at predicting the observed evolution of the spectral peak and the spectral mean. 
(v) The IS model, optimized over its free parameter, has the smallest overall error for three of the four experiments.  The vDysthe equation has the smallest overall error in the other experiment.

\section{Introduction}

\section{Introduction}
 
In a series of classic experiments, Lake {\emph{et al.}}~\cite{Lakeplus} and Lake \& Yuen~\cite{LY} examined the evolution of wave trains on deep water.  A paddle at one end of a narrow tank created a nearly two-dimensional, nearly monochromatic wave train with a particular frequency.  As the wave train traveled down the tank, it became modulated due to the growth of the Benjamin-Feir instability.  Further down the tank, the wave train returned to a nearly monochromatic form, however the dominant frequency was lower than the frequency created by the paddle.  This shift to a lower frequency in wave trains is referred to as frequency downshifting (FD).  Subsequent experiments, including those in Su {\emph{et al.}}~\cite{Suetal} and Melville~\cite{Melville}, demonstrated that the amplitude of the lower sideband grew and eventually overtook that of the carrier wave.  In 2005, Segur {\emph{et al.}}~\cite{sh}, conducted similar experiments and found that FD provided a means of distinguishing waves with ``small or moderate'' amplitudes, in which FD was not observed, and waves with ``large'' amplitudes, in which FD was observed.  In particular, they used successfully a nonlinear Schr\"odinger equation with linear damping to describe waves with small or moderate amplitudes.  Since that equation does not allow for FD, they considered waves that exhibited FD to have large amplitudes.

Many mathematical explanations for FD have been proposed.  Most, including Trulsen \& Dysthe~\cite{TD1990}, Hara \& Mei~\cite{Hara}, Kato \& Oikawa~\cite{Kato}, Islas-Schober~\cite{IslasSchober}, Brunetti \& Kasparian~\cite{Brunetti1}, and Brunetti {\emph{et al.}}~\cite{Brunetti2}, rely on wind and/or wave breaking as mechanisms for FD.  As there was no wind nor wave breaking in the Segur {\emph{et al.}}~\cite{sh} experiments, there must be another mechanism for FD.  A clear understanding of frequency downshifting and the mechanisms behind it will lead to a better understanding of the evolution of swell and how wave energy is transported across the ocean.  In this paper, we explore possible mathematical models/mechanisms for FD in the absence of wind and/or wave breaking using numerics and experiments.

Three quantities that are commonly used to quantify FD are the linear momentum,
\begin{equation}
    \mathcal{P}(\chi)=\frac{i}{2L}\int_0^L\big{(}BB_\xi^*-B_\xi B^*\big{)}d\xi,
    \label{P}
\end{equation}
the spectral peak, $\omega_p$, and the spectral mean, $\omega_m$.  Here $B$ is a dimensionless measure of the complex envelope of a slowly evolving, nearly monochromatic train of plane waves, $\chi$ is dimensionless distance down the tank, $\xi$ is dimensionless time, and $L$ is the dimensionless $\xi$-period of $B$.  (In this paper, all dimensional variables have overbars.  For example, $\bar{x}$ is the dimensional horizontal coordinate.)  The spectral peak is the frequency corresponding to the Fourier mode with largest magnitude.  The spectral mean is defined by the ratio
\begin{equation}
\omega_m=\frac{\mathcal{P}}{\mathcal{M}},
\end{equation}
where
\begin{equation}
\mathcal{M}(\chi)=\frac{1}{L}\int_0^L|B|^2d\xi.
\label{M}
\end{equation}
There is not a generally agreed upon definition for FD in the literature.  Some authors have defined FD to be a decrease in $\omega_m$, some have defined FD to be a decrease in $\omega_p$, and others have defined it to be a decrease in $\mathcal{P}$.   In hopes of reducing confusion, we use two definitions of FD.  We state that FD in the spectral mean sense occurs if $\omega_m$ decreases and that FD in the spectral peak sense occurs if $\omega_p$ decreases.  For example, Lake \& Yuen~\cite{LY} observed a decrease in $\omega_p$ (they did not measure $\omega_m$).  Segur {\emph{et al.}}~\cite{sh} observed that $\omega_p$ and $\omega_m$ decreased together.

The focus of this paper is to compare measurements of waves from four experiments with predictions from seven mathematical models that are in the literature.  One of the experiments exhibited frequency downshift while the other three did not.  The models include: the classical cubic nonlinear Schr\"odinger equation~\cite{Zak1968,SulemSulem} (NLS), the Dysthe equation~\cite{Dysthe} (Dysthe), the dissipative NLS equation~\cite{sh,DDZ} (dNLS), the viscous Dysthe equation~\cite{CarGov} (vDysthe), the Islas-Schober equation~\cite{IslasSchober} (IS), the Gordon equation~\cite{Gordon} (Gordon), and a new model presented herein, which we call the dissipative Gramstad-Trulsen equation (dGT), a dissipative generalization of the model derived by Gramstad-Trulsen~\cite{GT}.  We focus on FD and frequency upshift (FU) in both the spectral mean and spectral peak senses.  We show that the conservative models, NLS and Dysthe, do not accurately predict the measured spectral amplitudes.  We show that the Gordon model cannot accurately model FD in water wave experiments because it does not accurately model the evolution of $\mathcal{P}$ regardless of the choice of the free parameter.  We show that dNLS, vDysthe, dGT (none of which have free parameters), and IS (with appropriate choice of its free parameter) do the best at predicting the evolution of the spectral peak and mean.  In fact, with the optimal (defined below) parameter choice, IS typically provides the most accurate model of FD.

The remainder of this paper is organized as follows.  Section \ref{Equations} introduces the model equations and their properties.  Section \ref{Experiments} contains a description of the experimental facility and procedures.  Section \ref{Comparisons} contains results of the comparisons between experiments and predictions.  Finally, Section \ref{Summary} contains a summary and a list of some possible future work.

\section{Model equations}
\label{Equations}

Wu {\emph{et al.}}~\cite{WuLiuYue} proposed the following system for an infinitely-deep, weakly dissipative, two-dimensional fluid
\begin{subequations}
\begin{equation}
\bar{\phi}_{\bar{x}\bar{x}}+\bar{\phi}_{\bar{z}\bar{z}}=0, \hspace*{1cm}\mbox{for } -\infty<\bar{z}<\bar{\eta},
\end{equation}
\begin{equation}
    \bar{\phi}_{\bar{t}}+\frac{1}{2}\Big{(}\bar{\phi}_{\bar{x}}^2+\bar{\phi}_{\bar{z}}^2\Big{)}+\bar{g}\bar{\eta}=-\bar{\alpha}\bar{\phi}_{\bar{z}\bar{z}},\hspace*{1cm}\mbox{at } \bar{z}=\bar{\eta},
    \label{WLYb}
    \end{equation}
    \begin{equation}
    \bar{\phi}_{\bar{z}}=\bar{\eta}_{\bar{t}}+\bar{\eta}_{\bar{x}}\bar{\phi}_{\bar{x}},\hspace*{1cm}\mbox{at } \bar{z}=\bar{\eta},
    \label{WLYc}
    \end{equation}
    \begin{equation}
    |\nabla\bar{\phi}|\rightarrow0, \hspace*{1cm}\mbox{as } \bar{z}\rightarrow-\infty.
    \label{bottomBC}
    \end{equation}
\label{WLY}
\end{subequations}
Here $\bar{\phi}=\bar{\phi}(\bar{x},\bar{z},\bar{t})$ represents the velocity potential of the fluid, $\bar{\eta}=\bar{\eta}(\bar{x},\bar{t}) $ represents the free-surface displacement, $\bar{x}$ is the horizontal coordinate, $\bar{z}$ is the vertical coordinate, $\bar{t}$ is the temporal coordinate, $\bar{g}$ represents the acceleration due to gravity, and $\bar{\alpha}>0$ is a constant such that $\bar{\alpha}\bar{\phi}_{\bar{z}\bar{z}}$ represents dissipation from all sources.  The classical Euler equations are obtained from this system by setting $\bar{\alpha}=0$.  We note that Dias {\emph{et al.}}~\cite{DDZ} derived a version of (\ref{WLY}) that includes an additional (viscous/dissipative) term in equation (\ref{WLYc}) starting from a weakly viscous (linear) system.  However, that system does not conserve mass.

Wu {\emph{et al.}}~\cite{WuLiuYue} numerically integrated (\ref{WLY}) and compared with the Segur {\emph{et al.}}~\cite{sh} experimental results.  Wu {\emph{et al.}}~\cite{WuLiuYue} used the experimentally measured damping rate for $\bar{\alpha}$ that was obtained from the exponential decay of $\mathcal{M}$.  Their corresponding predictions from (\ref{WLY}) for the evolution of $\mathcal{P}$ and for the Fourier amplitudes of three frequency components of the spectrum agreed well with measurements.  

Four of the models for FD considered herein may be derived from (\ref{WLY}).  NLS and Dysthe can be derived when $\bar{\alpha}=0$ and dNLS and vDysthe can be derived when $\bar{\alpha}>0$.  These derivations are outlined below.  Two of the models, IS and Gordon, also follow the derivation (with $\bar{\alpha}=0$) but have additional, ad-hoc terms added, with free parameters.  The last model, dGT, is obtained from the Gramstad-Trulsen equation (GT)~\cite{GT} by adding three ad-hoc terms (but no free parameters) that allow for dissipation. The GT equation can be derived from \ref{WLY} with $\bar{\alpha}=0$.  All of these models are discussed in more detail below.

\subsection{Asymptotic models}

In order to model the evolution of a slowly modulated wave train, assume
\begin{subequations}
    \begin{equation}
        \bar{\eta}(\bar{x},\bar{t},\bar{X},\bar{T})=\epsilon^3\bar{H}(\bar{X},\bar{T})+\epsilon \bar{B}(\bar{X},\bar{T})\mbox{e}^{i\bar{\omega}_0\bar{t}-i\bar{k}_0\bar{x}}+\epsilon^2\bar{B}_2(\bar{X},\bar{T})\mbox{e}^{2(i\bar{\omega}_0\bar{t}-i\bar{k}_0\bar{x})}
        +\dots+c.c.,
        \label{eta}
        \end{equation}
\begin{equation}
    \begin{split}
    \bar{\phi}(\bar{x},\bar{z},\bar{t},\bar{X},\bar{Z},\bar{T})=\epsilon^2\bar{\Phi}(\bar{X},\bar{Z},\bar{T})&+\epsilon
    \bar{A}_1(\bar{X},\bar{Z},\bar{T})\mbox{e}^{\bar{k}_0\bar{z}+i\bar{\omega}_0\bar{t}-i\bar{k}_0\bar{x}}\\ &+\epsilon^2\bar{A}_2(\bar{X},\bar{Z},\bar{T})\mbox{e}^{2(\bar{k}_0\bar{z}+i\bar{\omega}_0\bar{t}-i\bar{k}_0\bar{x})}
    +\dots+c.c.,
    \end{split}
\end{equation}
\label{ansatz}
\end{subequations}
where $\bar{\omega}_0\in\mathbb{R}$ and $\bar{k}_0>0$ are constants representing the frequency and wavenumber of the carrier wave respectively, $\epsilon=2\bar{a}_0\bar{k}_0\ll 1$ is the dimensionless wave steepness, $\bar{a}_0$ represents a typical wave amplitude, and $c.c.$ stands for complex conjugate.  The slow, dimensional variables are defined by $\bar{X}=\epsilon\bar{x}$, $\bar{Z}=\epsilon\bar{z}$, and $\bar{T}=\epsilon\bar{t}$.  We assume that dissipative effects are small by setting $\bar{\alpha}=\epsilon^2\tilde{\alpha}$.  Note that the sign in the exponentials in equation (\ref{ansatz}) is different than the choice typically used in the literature.  This choice is made so that a decrease in the experimental frequency corresponds to a decrease in the frequency in equations (\ref{WLY})-(\ref{ansatz}) and in the asymptotic models introduced below.  Finally, in order to compare with our unidirectional experiments, we only consider waves that travel to the right and therefore require $\bar{\omega}_0>0$.

Substituting equation (\ref{ansatz}) into (\ref{WLY}) gives the deep-water linear dispersion relationship
\begin{equation}
\bar{\omega}_0^2=\bar{g}\bar{k}_0,
\end{equation}
at leading order.  Next, nondimensionalize and transform to enter a coordinate frame moving with the linear group velocity, $\frac{\bar{\omega}_0}{2\bar{k}_0}$, by introducing the change of variables
\begin{subequations}
\begin{equation}
\xi=\bar{\omega}_0\bar{T}-2\bar{k}_0\bar{X},
\label{xiCOV}
\end{equation}
\begin{equation}{}
\chi=\epsilon\bar{k}_0\bar{X},
\end{equation}
\begin{equation}
\bar{k}_0\bar{B}(\bar{X},\bar{T})=B(\xi,\chi),
\end{equation}
\begin{equation}
\frac{\bar{k}_0^2}{\bar{\omega}_0}\tilde{\alpha}=\delta,
\end{equation}
\label{COV}
\end{subequations}
where $\xi$ is nondimensional time, $\chi$ is nondimensional distance down the tank, $B$ is the dimensionless complex amplitude of the envelope, and $\delta$ is the nondimensional dissipation parameter.  The vDysthe equation,
\begin{equation}
iB_\chi+B_{\xi\xi}+4|B|^2B+i\delta
B+\epsilon\Big{(}-8iB^2B^*_\xi-32i|B|^2B_\xi-8\Big{(}\mathcal{H}\big{(}|B|^2\big{)}\Big{)}_\xi B+5\delta B_\xi\Big{)}=0,
\label{vDysthe}
\end{equation}
arises as a solvability condition at $\mathcal{O}(\epsilon^4)$.  Here $\mathcal{H}$ is the Hilbert transform defined by
\begin{equation}
\mathcal{H}\big{(}f(\xi)\big{)}=\sum_{k=-\infty}^{\infty}-i\,\mbox{sgn}(k)\hat{f}(k)\mbox{e}^{2\pi ik\xi/L},
\end{equation}
and the Fourier transform of a function $f(x)$ is defined by
\begin{equation}
\hat{f}(k)=\frac{1}{L}\int_0^L f(\xi)\mbox{e}^{-2\pi ik\xi/L}d\xi,
\end{equation}
where $L$ is the $\xi$-period of the experimental time series.  Carter \& Govan~\cite{CarGov} showed that the vDysthe equation accurately models data from experiments in which frequency downshift was observed and experiments in which frequency downshift was not observed.  Kimmoun {\emph{et al.}}~\cite{vDystheBigger} showed that the vDysthe equation accurately models data from experiments conducted in a much larger experimental tank.

When $\delta=\epsilon=0$, the viscous Dysthe equation reduces to the NLS equation, 
\begin{equation}
    iB_\chi+B_{\xi\xi}+4|B|^2B=0.
    \label{NLS}
\end{equation}
The NLS equation was first derived as a model of water waves by Zakharov~\cite{Zak1968}.  It has been well studied; see for example Ablowitz \& Segur~\cite{AS} and Sulem \& Sulem~\cite{SulemSulem}.  

When $\epsilon=0$, the vDysthe equation reduces to the dNLS equation,
\begin{equation}
    iB_\chi+B_{\xi\xi}+4|B|^2B+i\delta B=0.
    \label{dNLS}
    \end{equation}
This equation has been been shown to compare well with experiments in which FD did not occur; see for example Segur {\emph{et al.}}~\cite{sh} and Wu {\emph{et al.}}~\cite{WuLiuYue}.

When $\delta=0$, the vDysthe equation reduces to the Dysthe~\cite{Dysthe} equation, which is also known as the modified NLS equation,
\begin{equation}
    iB_\chi+B_{\xi\xi}+4|B|^2B+\epsilon\Big{(}-8iB^2B^*_\xi-32i|B|^2B_\xi-8\Big{(}\mathcal{H}\big{(}|B|^2\big{)}\Big{)}_\xi B\Big{)}=0.
    \label{Dysthe}
    \end{equation}
Lo \& Mei~\cite{LoMei} compared numerical simulations with experimental measurements and showed that the Dysthe equation more accurately predicts the evolution of wave trains than does the NLS equation.

In 2011, Gramstad-Trulsen~\cite{GT} derived a Hamiltonian fourth-order NLS-type equation (in other words, a Hamiltonian version of the Dysthe equation).  Islas-Schober~\cite{IslasSchober} proposed the following dissipative generalization of this equation
\begin{equation}
iB_\chi+B_{\xi\xi}+4|B|^2B+i\delta B+\epsilon\Big{(}-32i|B|^2B_\xi+(-8+i\beta_1)\Big{(}\mathcal{H}\big{(}|B|^2\big{)}\Big{)}_\xi B\Big{)}=0,
\label{IS}
\end{equation}
where $\beta_1$ is an arbitrary constant.  We refer to this equation as the IS equation.  The constant $\beta_1$ must be positive in order for that term to represent additional dissipative effects (see equation (\ref{ISMchi})). 

We propose a different generalization of the Gramstad-Trulsen equation
\begin{equation}
iB_\chi+B_{\xi\xi}+4|B|^2B+i\delta
B+\epsilon\Big{(}-32i|B|^2B_\xi-8\Big{(}\mathcal{H}\big{(}|B|^2\big{)}\Big{)}_\xi B+5\delta B_\xi\Big{)}-10i\epsilon^2\delta B_{\xi\xi}=0,
\label{gvDysthe}
\end{equation}
as a model for the evolution of wave trains on deep water.  The three terms including $\delta$ come from extending the derivation of the vDysthe equation one order higher.  We refer to this equation as the dissipative Gramstad-Trulsen (dGT) equation.  This equation avoids some of the difficulties exhibited by the Dysthe and vDysthe equations as well as the free parameter of IS.  These difficulties are discussed along with the properties of these equations in the next section.

Frequency downshift has also been observed in experiments on electromagnetic waves in optical fibers~\cite{MM}.  Gordon~\cite{Gordon} derived the following equation for the approximate evolution of the complex amplitude of the slowly varying envelope of the carrier wave by including Raman effects
\begin{equation}
iB_\chi+B_{\xi\xi}+4|B|^2B+\epsilon\beta_2\big{(}|B|^2\big{)}_\xi B=0,
\label{Gordon}
\end{equation}
where $\beta_2>0$ is an arbitrary constant in the water waves setting.  We refer to this equation as the Gordon equation.  Although electromagnetic waves in optical fibers and surface water waves are quite different physically, both can be accurately modeled using NLS-type equations.  It is reasonable to compare (\ref{Gordon}) with data to determine whether or not a $\beta_2$ exists for water waves.  Therefore, we include this optics model of FD in our study of FD in water waves.  (Though, in Section \ref{Comparisons}, we show that no such constant exists for our water wave experiments.)

\subsection{Model properties}
\label{propertiessection}

Here we analyze the models to determine, if possible, whether their solutions may exhibit  FD in the spectral mean sense, and whether their solutions preserve $\mathcal{M}$ and/or $\mathcal{P}$.

\subsubsection{Predicted evolution of $\omega_m$, $\mathcal{M}$, and $\mathcal{P}$}

The conservative (non-dissipative) asymptotic models for the evolution of wave trains on deep water, NLS, Dysthe, and GT are derived from Euler's equations, (\ref{WLY}) with $\bar{\alpha}=0$.  Euler's equations preserve $\mathcal{M}$, $\mathcal{P}$, and $\omega_m$.  FD is a phenomenon that breaks this structure of the system.  In particular, FD requires for at least two of $\mathcal{M}$, $\mathcal{P}$, and $\omega_m$ to not be conserved.  Below we consider how $\mathcal{M}$, $\mathcal{P}$, and $\omega_m$ evolve according to the various approximate models.  From these results, we determine whether or not each model predicts FD in the spectral mean sense. It is more difficult to obtain analytic results about FD in the spectral peak sense.  However, Segur {\emph{et al.}}~\cite{sh} noted that in their experiments, the FD in the spectral peak sense was always accompanied by a decrease in $\mathcal{P}$.  Thus, our conclusions for FD in the spectral peak sense are based on the evolution of $\mathcal{P}$.  These conclusions are consistent with our numerical results that are discussed in Section \ref{Comparisons}.  We also note that in experiments, there is damping, which causes a nearly exponential decay of both $\mathcal{M}$ and $\mathcal{P}$.  Thus, we seek a model that predicts both this decay as well as the FD exhibited by an evolving $\mathcal{P}$.  The vDysthe equation, (\ref{vDysthe}), the IS equation, (\ref{IS}), and the dGT equation, (\ref{gvDysthe}), have all of these ingredients.

The vDysthe equation does not preserve $\mathcal{M}$, $\mathcal{P}$, or $\omega_m$ in $\chi$ (dimensionless distance down the tank).   The $\chi$ dependencies of these quantities are given by
\begin{subequations}
\begin{equation}
\frac{d\mathcal{M}}{d\chi}=-2\delta\mathcal{M}-10\epsilon\delta\mathcal{P},
\label{Mchi}
\end{equation}{}
\begin{equation}
\frac{d\mathcal{P}}{d\chi}=-2\delta\mathcal{P}-10\epsilon\delta\mathcal{Q}-16\epsilon\mathcal{R},
\label{Pchi}
\end{equation}
\begin{equation}
\frac{d\omega_m}{d\chi}=\frac{d}{d\chi}\Big{(}\frac{\mathcal{P}}{\mathcal{M}}\Big{)}=-\frac{10\epsilon\delta}{\mathcal{M}^2}\Big{(}\mathcal{MQ}-\mathcal{P}^2\Big{)}-16\epsilon\frac{\mathcal{R}}{\mathcal{M}},
\label{omegamchi}
\end{equation}
\label{MPchi}
\end{subequations}
where
\begin{subequations}
\begin{equation}
\mathcal{Q}=\frac{1}{L}\int_0^L|B_\xi|^2d\xi,
\end{equation}
\begin{equation}
\mathcal{R}=\Im\Big{(}\frac{1}{L}\int_0^L |B|^2B_{\xi\xi}B^* d\xi\Big{)}.
\end{equation}
\end{subequations}

Equation (\ref{Mchi}) establishes that to leading order, $\mathcal{M}$ decays exponentially in $\chi$.  This allows $\delta$ to be determined empirically (see Section \ref{Experiments}).  Equation (\ref{Mchi}) also establishes that $\mathcal{M}$ decreases more rapidly when $\mathcal{P}>0$ and more slowly when $\mathcal{P}<0$.  This suggests a preference for waves with negative wavenumbers (i.e.~$\mathcal{P}<0$).  The first term on the right-hand side of equation (\ref{Pchi}) establishes that to leading order, $\mathcal{P}$ decays exponentially in $\chi$.  The second term provides a preference for downshifting (over upshifting) in the spectral peak sense, because $\mathcal{Q}$ is positive for all $B$ that are not constant in $\xi$.  This term competes with the term including $\mathcal{R}$, which could either enhance downshifting or cause upshifting, because its sign is indefinite.  In (\ref{omegamchi}), the first term on the right-hand side provides a preference for downshifting (over upshifting) in the spectral mean sense, because the Cauchy-Schwarz inequality establishes that $(\mathcal{MQ}-\mathcal{P})\ge0$. Again, the $\mathcal{R}$ term could either enhance the downshifting or cause upshifting.  We note that Ma {\emph{et al.}}~\cite{Ma} observed upshifting in the spectral peak sense in some of their experiments, so a good model for downshifting should allow for the possibility of both upshifting and downshifting in both the spectral peak and the spectral mean.

Setting $\delta=\epsilon=0$ in equation (\ref{MPchi}) shows that the NLS equation preserves $\mathcal{M}$, $\mathcal{P}$, and $\omega_m$.  Therefore NLS cannot model frequency downshift in the spectral mean sense.  Setting $\epsilon=0$ in equation (\ref{MPchi}) establishes that although the dNLS equation does not preserve $\mathcal{M}$ nor $\mathcal{P}$, it preserves $\omega_m$ and therefore cannot model FD in the spectral mean sense either.  Setting $\delta=0$ in (\ref{MPchi}), but allowing $\epsilon>0$, recovers the results from the (inviscid) Dysthe equation.  Equation (\ref{omegamchi}) shows that the (inviscid) Dysthe model predicts either upshifting or downshifting in the spectral mean sense.  There is no preference for downshifting, which is usually (but not always) observed. We note, however, that the (inviscid) Dysthe model predicts $\mathcal{M}$ to be a constant, which does not agree with experimental observations. It also does not capture the exponential decay of $\mathcal{P}$ observed in small-amplitude experiments.

The IS equation does not preserve $\mathcal{M}$, $\mathcal{P}$, or $\omega_m$.  The $\chi$ dependencies of these quantities are given by
\begin{subequations}
\begin{equation}
\frac{d\mathcal{M}}{d\chi}=-2\delta\mathcal{M}-\frac{2\epsilon\beta_1}{L}\int_0^L|B|^2\Big{(}\mathcal{H}\big{(}|B|^2\big{)}\Big{)}_\xi d\xi,
\label{ISMchi}
\end{equation}{}
\begin{equation}
\frac{d\mathcal{P}}{d\chi}=-2\delta\mathcal{P}--\frac{i\epsilon\beta_1}{L}\int_0^L\Big{(}BB_\xi^*-B_\xi B^*\Big{)}\Big{(}\mathcal{H}\big{(}|B|^2\big{)}\Big{)}_\xi d\xi,
\label{ISPchi}
\end{equation}
\begin{equation}
\frac{d\omega_m}{d\chi}=-\frac{2\epsilon\beta_1}{\mathcal{M}^2}\Bigg{(}\frac{i\mathcal{M}}{2L}\int_0^L\big{(}BB_\xi^*-B_\xi B^*\big{)}\Big{(}\mathcal{H}\big{(}|B|^2\big{)}\Big{)}_\xi~d\xi-\frac{\mathcal{P}}{L}\int_0^L|B|^2\Big{(}\mathcal{H}\big{(}|B|^2\big{)}\Big{)}_\xi~d\xi\Bigg{)}.
\label{ISomegam}
\end{equation}
\label{ISMPchi}
\end{subequations}
Since $\beta_1$ is an arbitrary constant, the IS model allows for both FD and FU.

The Gordon equation preserves $\mathcal{M}$, but neither $\mathcal{P}$, nor $\omega_m$.  The $\chi$ dependencies of these quantities are given by
\begin{subequations}
\begin{equation}
\frac{d\mathcal{P}}{d\chi}=-\epsilon\beta_2\mathcal{S},
\label{GordonPchi}
\end{equation}
\begin{equation}
\frac{d\omega_m}{d\chi}=-\epsilon\beta_2\frac{\mathcal{S}}{\mathcal{M}},
\label{Gordonomegamchi}
\end{equation}
\end{subequations}
where
\begin{equation}
\mathcal{S}=\frac{1}{L}\int_0^L \Big{(}\big{(}|B|^2\big{)}_\xi\Big{)}^2d\xi.
\label{Sdef}
\end{equation}
Equation (\ref{Gordonomegamchi}) establishes that the Gordon equation, with $\beta_2>0$, predicts FD in the spectral mean sense for all $B$ except those that have constant modulus in $\xi$.

The dGT equation does not preserve $\mathcal{M}$, $\mathcal{P}$, or $\omega_m$.  The $\chi$ dependencies of these quantities are given by
\begin{subequations}
\begin{equation}
\frac{d\mathcal{M}}{d\chi}=-2\delta\mathcal{M}-10\epsilon\delta\mathcal{P}-20\epsilon^2\delta\mathcal{Q},
\label{dGTMchi}
\end{equation}
\begin{equation}
\frac{d\mathcal{P}}{d\chi}=-2\delta\mathcal{P}-10\epsilon\delta\mathcal{Q}-20\epsilon^2\delta\mathcal{T},
\label{dGTPchi}
\end{equation}
\begin{equation}
\frac{d\omega_m}{d\chi}=\frac{-10\epsilon\delta\big{(}\mathcal{MQ}-\mathcal{P}^2\big{)}-20\epsilon^2\delta\big{(}\mathcal{MT}-\mathcal{QP}\big{)}}{\mathcal{M}^2},
\label{dGTomegamchi}
\end{equation}
\label{dGTMPchi}
\end{subequations} 
where
\begin{equation}
\mathcal{T}=\Im\Big{(}\frac{1}{L}\int_0^L B_{\xi\xi}B_{\xi}^*d\xi\Big{)}.
\end{equation}
Since $\mathcal{M}\ge0$ and $\mathcal{Q}\ge0$, equation (\ref{dGTMchi}) establishes that $\mathcal{M}$ decreases more rapidly when $\mathcal{P}>0$.  This suggests a preference for waves with negative wavenumbers.  The Cauchy-Schwarz inequality establishes that $(\mathcal{MQ}-\mathcal{P}^2)\ge0$.  The Cauchy-Schwarz inequality also establishes that if $\mathcal{P}>0$, then $(\mathcal{MT}-\mathcal{QP})\ge0$ or if $\mathcal{P}<0$, then $(\mathcal{MT}-\mathcal{QP})\le0$.  Therefore, equation (\ref{dGTomegamchi}) establishes that under the dGT equation, the spectral mean will not increase as long as the $\mathcal{O}(\epsilon^2)$ term is smaller than the $\mathcal{O}(\epsilon)$ term, as it should be.  However, we show below (see Section \ref{Comparisons}) that in one of the experiments herein (Expt D), an increase in the spectral mean is observed.

Another important property of the dGT equation is that all solutions to the linearized dGT equation decay to zero in $\chi$ if $\delta>0.$  The linearized vDysthe equation does not have this property.  Solutions of the linearized vDysthe equation with wavenumber less than $-\frac{1}{5\epsilon}$ grow exponentially in $\chi$ when $\delta>0$.  Wavenumbers in this range are outside of the narrow-bandwidth region where the equation is meant to be valid; nevertheless, it is a limitation of the model.

\subsubsection{Discussion of model properties}

The models considered herein are generalizations of NLS that contain additional nonlinear, broad-bandedness, and dissipative terms.  There are two things we are seeking.  One is to determine which effects can lead to frequency down/upshifting, in general.  The other is to find the mechanisms that agree with experimental measurements.

In order to get down/upshifting in general, the symmetry that is inherent to NLS must be broken.  The Dysthe model breaks this symmetry because there is a nonlinear term with a single $\xi$ derivative.  The impact of this term on frequency down/upshifting depends on the initial conditions and other studies have shown that it predicts both temporary FD and temporary FU.  However, predictions obtained from the Dysthe equation do not agree with experiments because it does not include damping.

Another mechanism that breaks the symmetry and gives down/upshifting is nonlinear viscous effects such as the additional terms considered by Islas \& Schober\cite{IslasSchober}.  It is currently unknown how these terms would arise from the physical system.  Regardless, this approach is encouraging because we were able to find a free parameter for most experiments such that the model agrees reasonably well with the experiments.  One can conjecture (or at least hope) that if one derived a high-order NLS-like equation from the full Navier-Stokes equations with viscous boundary conditions, one would derive this term.

The vDysthe equation breaks the symmetry by including a term that arises from linear dissipation that includes a single $\xi$ derivative.  This model is also encouraging, because it agrees with most experiments.

All three effects seem to be important: broad-bandedness, nonlinearity, and viscosity, because they interact. Broad-bandedness and nonlinearity can work against each other - broad-bandedness tends to stabilize while nonlinearity tends to destabilize (as represented by the BFI).  And for viscous effects to be important for downshifting, there has to be some threshold of nonlinearity.

\section{Experimental apparatus and results}
\label{Experiments}

To test the accuracy of the models, we compare predictions of Fourier amplitudes $\mathcal{M}$, $\mathcal{P}$,  $\omega_p$, and $\omega_m$, from the models, with data already in the literature and with two new experiments.  Experiments A and B were presented in Segur {\emph{et al.}}~\cite{sh} and reconsidered in Carter \& Govan~\cite{CarGov}.  Experiments C and D are new.  All four experiments were conducted in the William G.~Pritchard Fluid Mechanics Laboratory in the Mathematics Department at Penn State. The procedures for experiments A and B are described in Segur {\emph{et al.}}~\cite{sh}.  The procedures for experiments C and D are very similar, although the equipment involved was different. The wave channel used for A and B was 43~ft long.
The wave channel used for C and D was 50~ft long. Both channels had glass bottoms and sidewalls and were 10~in wide.  
The tank walls were cleaned with alcohol; then water was added.  The air-water interface was cleaned by skimming (A and B) or blowing (C and D) the interfacial layer to one end of the tank where it was vacuumed -- the resulting still water depth for all experiments was $20$~cm.  Waves were generated in all four experiments with anodized, wedge-shaped plungers that spanned the width of the tank and were oscillated vertically using feed-back, programmable control.  The cross-section of the wedge was exponential for A and B (with a fall-off that corresponds to the velocity field for a 3.33 Hz wave) and was triangular for C and D (with a slope corresponding to a linear approximation to the exponential of the paddle for A and B).  For Experiments A and B, which used PMAC - Delta Tau Data Systems for motion control,  the wedge was oscillated with a time series given by
$\bar{\eta}_p(t) =  \bar{a}_f \sin(\bar{\omega}_{0f} t)\Big(1 + r \sin(\bar{\omega}_{1f} \bar{t})\Big)$, where $\bar{a}_f$ is the forcing amplitude of the carrier wave, $\bar{\omega}_{0f}$ is the carrier wave frequency, $r$ is the ratio of perturbation amplitude to $\bar{a}_f$, and $\bar{\omega}_{1f}$ is the perturbation frequency.  Expts C and D used ARCS software for motion control. For C, the wedge was oscillated with a time series given by
$\bar{\eta}_p(t) =  \bar{a}_f  \sin(\bar{\omega}_{0f} \bar{t} ) \Big(1 -  r \cos(\bar{\omega}_{1f} \bar{t})\Big)$,
and for Experiment D, $\bar{\eta}_p(t) =  \bar{a}_f  \sin(\bar{\omega}_{0f} \bar{t} ) + \bar{a}_f r \sin\Big((\bar{\omega}_{0f} + \bar{\omega}_{1f})\bar{t}\Big)$.  The forcing for Experiment D is different so as to purposefully force the upper sideband but not the lower sideband.  The forcing amplitudes and frequencies are given in Table~{\ref{WavemakerTable}}.

\begin{table}
\begin{center}
\begin{tabular}{|c|c|c|c|c|c|c|c|c|c|c|}
\hline 
parameter & Expt A & Expt B & Expt C & Expt D\\
\hline
$\bar{\omega}_{0f}/(2\pi)$~~~(Hz) & 3.33 & 3.33 & 3.33 & 3.33\\
$\bar{\omega}_{1f}/(2\pi)$~~~(Hz) & 0.17 & 0.17 & 0.11 & 0.11\\
$\bar{a}_f$~~(cm) & 0.25 & 0.25 & 0.50 & 0.50\\
$r$ & 0.14 & 0.33 & 0.50 & 0.50\\ 
\hline
\end{tabular}
\end{center}
\caption {Wave maker parameters: carrier wave frequency, $\bar{\omega}_{0f}$; modulation frequency, $\bar{\omega}_{1f}$; forcing amplitude of carrier wave, $\bar{a}_f$;  ratio of forcing amplitudes of the modulation to that of the carrier wave, $r$. In Expt.~D, only the upper sideband was forced.}
\label{WavemakerTable}
\end{table}

\begin{figure}
\centering
\includegraphics[width=12cm]{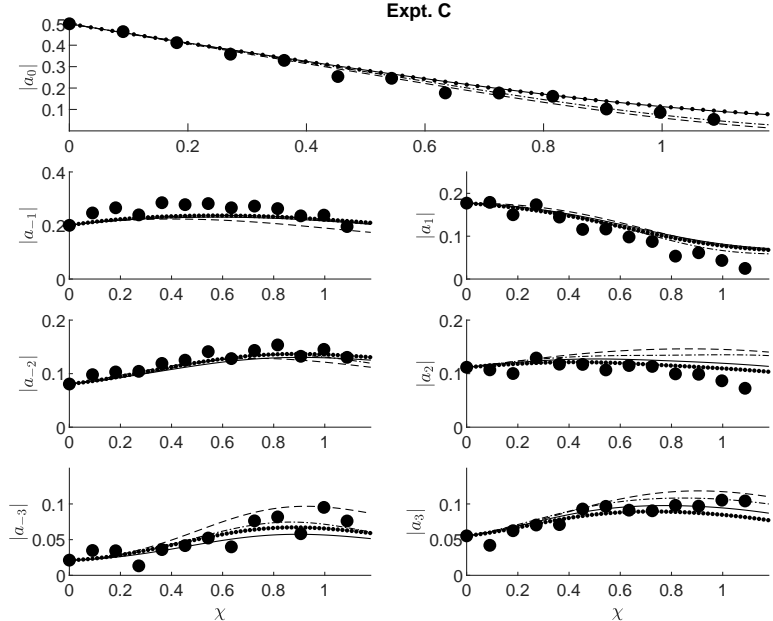}
\caption{Plots of the nondimensional amplitudes of the carrier wave, $|a_0|$, and the three most energetic sideband pairs, $|a_{\pm j}|$, versus nondimensional distance down the tank, $\chi$, for Expt C.  The large dots correspond to experimental measurements, the dashed curves correspond to dNLS, the solid curves correspond to vDysthe, the dash-dot curves correspond to optimized IS, and the dotted curves correspond to dGT.}
\label{FAmpsFeb}
\end{figure}

\begin{figure}
\centering
\includegraphics[width=12cm]{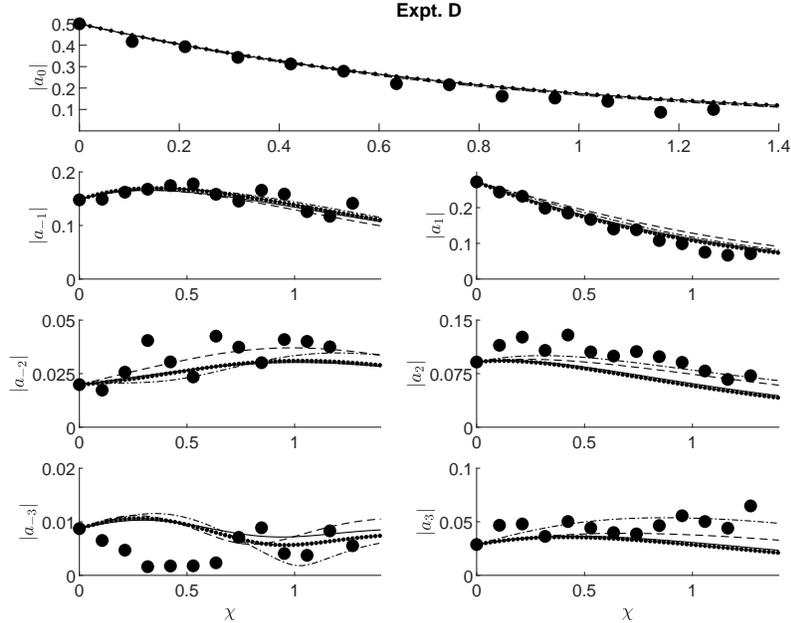}
\caption{Plots of the nondimensional amplitudes of the carrier wave, $|a_0|$, and the three most energetic sideband pairs, $|a_{\pm j}|$, versus nondimensional distance down, $\chi$, the tank for Expt D.  The large dots correspond to experimental measurements, the dashed curves correspond to dNLS, the solid curves correspond to vDysthe, the dash-dot curves correspond to optimized IS, and the dotted curves correspond to dGT.}
\label{FAmpsMay}
\end{figure}

\begin{figure}
\centering
\includegraphics[width=16cm]{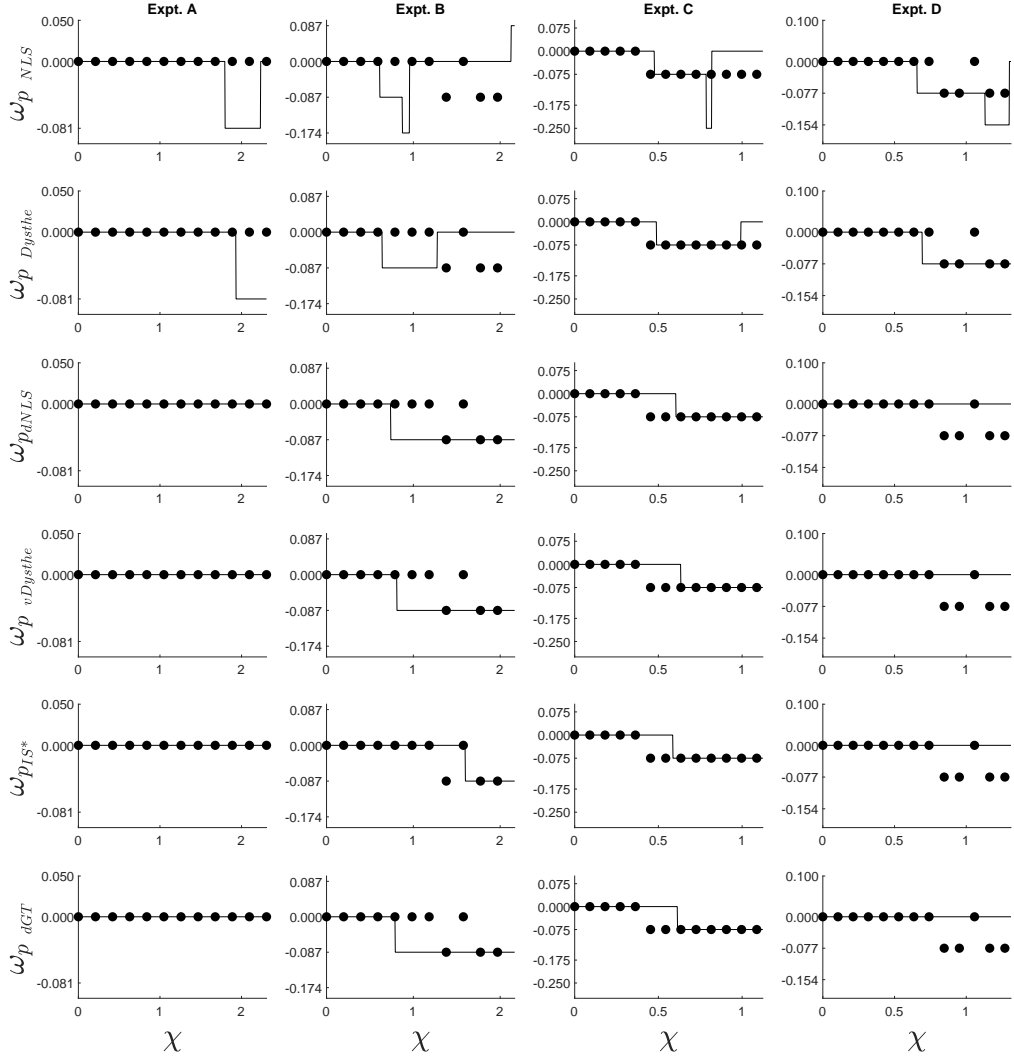}
\caption{Plots of the spectral peak, $\omega_p$, versus dimensionless distance down the tank, $\chi$.  The columns show results for the four experiments.  The rows correspond to the results from NLS, Dysthe, dNLS, vDysthe, IS$^*$, and dGT.  The dots correspond to experimental measurements and the lines correspond to the model predictions.}
\label{Omegaps}
\end{figure}

\begin{figure}
\centering
\includegraphics[width=16cm]{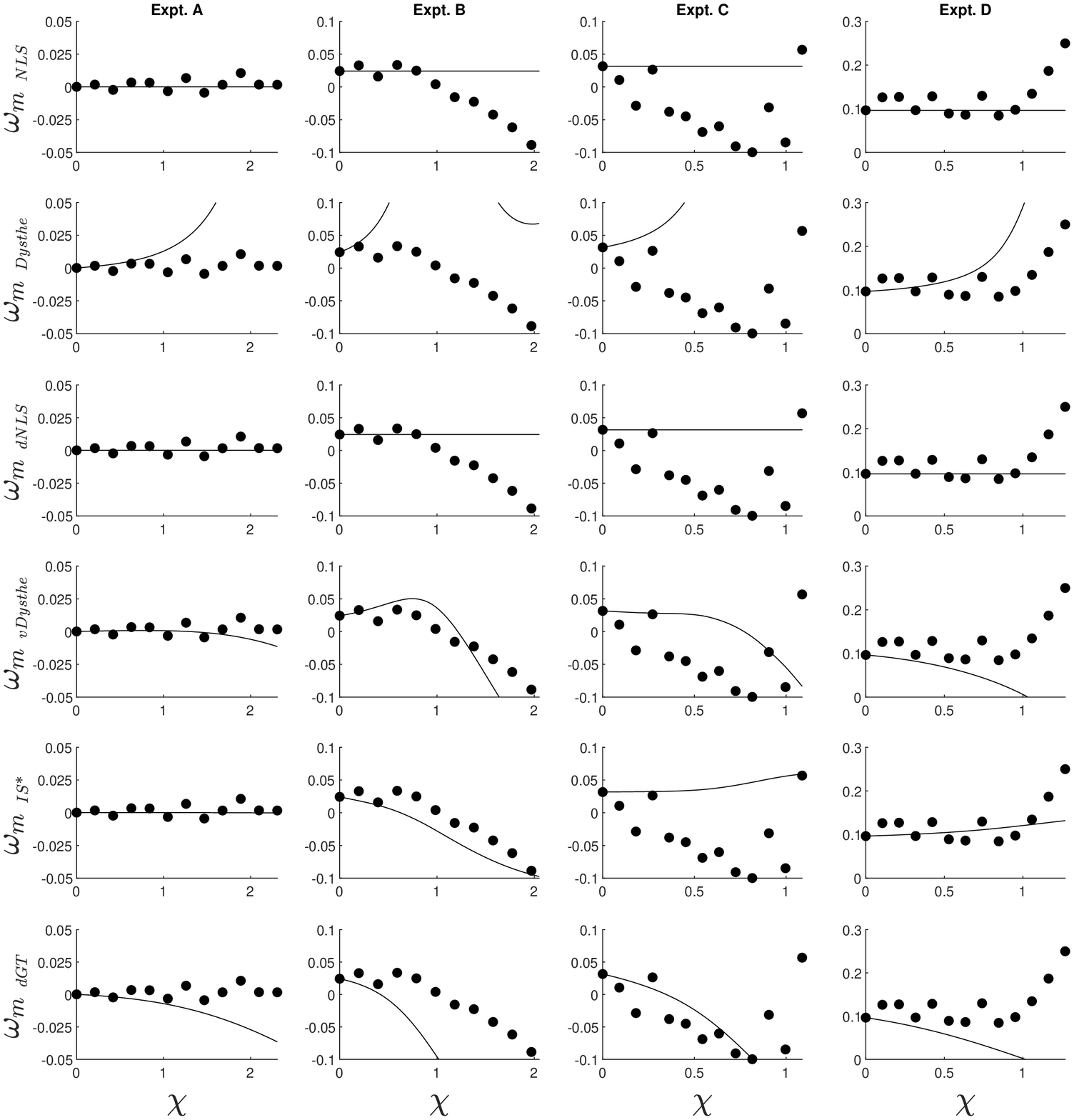}
\caption{Plots of the spectral mean, $\omega_m$, versus dimensionless distance down the tank, $\chi$.  The columns show results for the four experiments.  The rows correspond to the results from NLS, Dysthe, dNLS, vDysthe, IS$^*$, and dGT.  The dots correspond to experimental measurements and the curves correspond to the model predictions.}
\label{Omegams}
\end{figure}

Capacitance-type wave gauges whose signals were imported into LabVIEW were used to measure time series of surface displacement at a point. For Experiments A and B, the gauge was a non-intrusive capacitance-type gauge that spanned 12.7~cm of the width of the tank and 6~mm in the direction of wave propagation. For Expts C and D, the gauge was an intrusive glass tube, 1.6~mm in outside diameter, which contained a conductor and was sealed at the underwater end.  Each of the four experiments consisted of a set of 10-13 experiments with the gauges located $x_m=128+50(m-1)$ for $m=1,2,\dots,M$ centimeters from the wave maker.  The values of $M$ for the four experiments are included in Table \ref{DataTable}, which contains all of the experimentally measured parameters.  
The Fourier transforms of the time series were obtained to determine the complex amplitudes of the carrier wave and its sidebands.   Their magnitudes were then compared with predictions from numerical computations of the models - the comparisons are shown in Figures \ref{FAmpsFeb}, \ref{FAmpsMay}, \ref{Omegaps}, and \ref{Omegams} and are discussed in Section \ref{Comparisons}.  Their magnitudes are compared with predictions from numerical computations of the models.  We note that both the amplitude and phase of the measurements are required to initialize the numerical computations done for these comparisons.  The integrals for $\mathcal{M}$ and $\mathcal{P}$ are obtained using Parseval's theorem, the measured Fourier coefficients at each measurement location, and the notion that the derivatives in $\mathcal{P}$ correspond to multiplication by the corresponding difference frequencies.

\begin{table}
\begin{center}
\begin{tabular}{|c| c c c c |}
\hline
parameter & Expt A & Expt B & Expt C & Expt D \\
\hline
$M$ & 12 & 11 & 13 & 13 \\
$\bar{k}_0$~~~(cm$^{-1}$) & 0.4478 & 0.4470 & 0.4472 & 0.4472 \\
$\epsilon$ & $9.677*10^{-2}$ & $9.388*10^{-2}$ & $6.366*10^{-2}$ & $6.878*10^{-2}$ \\
$\delta$ & 0.2640 & 0.3223 & 0.5021 & 0.8129 \\
$N$ & 41 & 39 & 50 & 45 \\
$\bar{t}_f$~~~(sec) & 24.28 & 23.40 & 30.00 & 27.00\\
BFI  & 3.919 & 3.661 & 4.244 & 4.127 \\
\hline
\end{tabular}
\caption{Experimentally measured parameters for each of the four experiments.}
\label{DataTable}
\end{center}
\end{table}

\begin{figure}
\centering
\includegraphics[width=16cm]{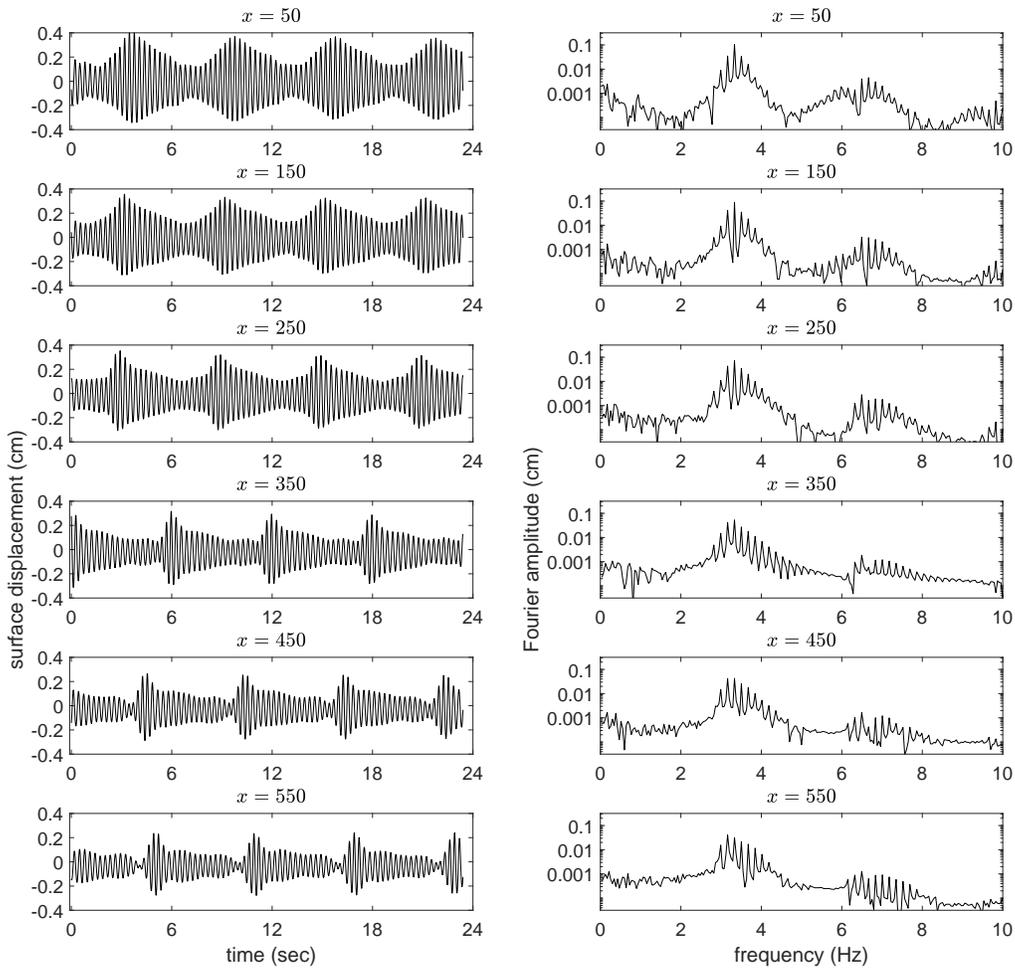}
\caption{Plots from every other gauge for Expt B.  The first column contains plots of surface displacement (in cm) versus time (in sec).  The second column contains plots of the magnitudes of the corresponding Fourier coefficients (in cm) versus frequency (in Hz).}
\label{TimeSeries}
\end{figure}

Figure \ref{TimeSeries} contains plots of the surface displacement (in cm) at fixed distances ($x+128$~cm) from the wave maker as functions of time (in sec) and the corresponding Fourier magnitudes (in cm) versus frequency (in Hertz) for Experiment B.  For conciseness, only the results from every other gauge are included.  Plots from the other experiments are similar.  The plots for Expt B show that FD in the spectral peak sense occurs between the gauges at $x=350$ and $x=450$ because the spectral peak decreases from $3.33$~Hz (the carrier wave) to $3.16$~Hz (the first lower sideband) between these gauges.  

\begin{figure}
    \centering
    \includegraphics[width=10cm]{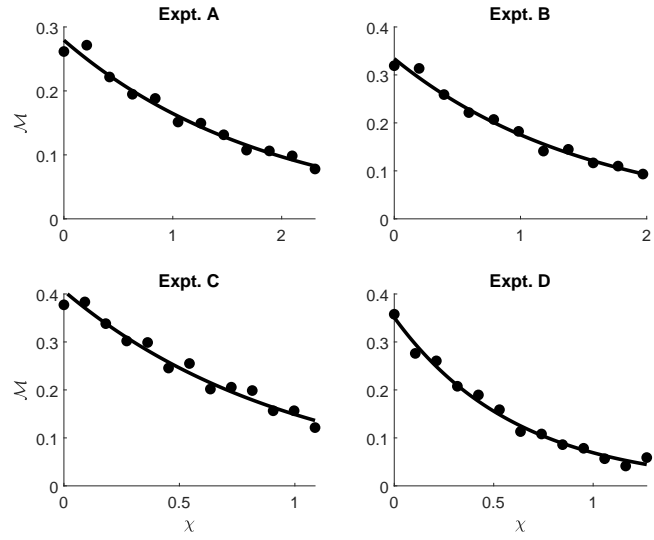}
    \caption{Plots of dimensionless mass, $\mathcal{M}$, versus dimensionless distance down the tank, $\chi$, for the four experiments.  The dots correspond to experimental measurements and the curves are the best-fit exponentials.}
    \label{Mnorms}
\end{figure}

\begin{figure}
    \centering
    \includegraphics[width=10cm]{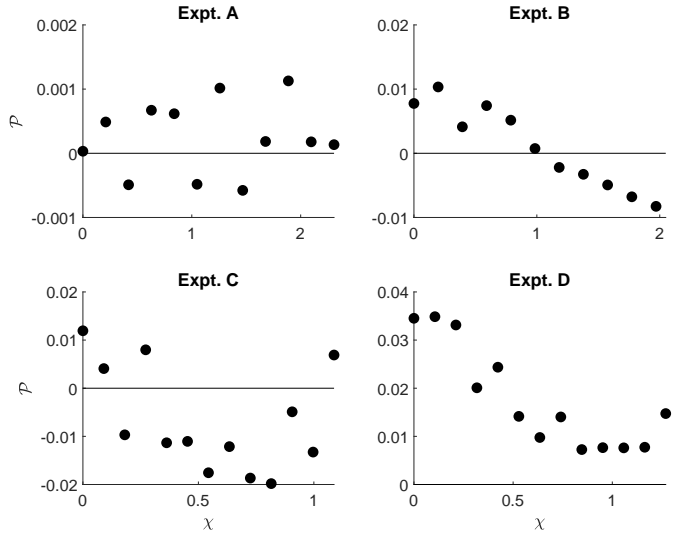}
    \caption{Plots of dimensionless linear momentum, $\mathcal{P}$, versus dimensionless distance down the tank, $\chi$, for the four experiments.  The horizontal lines correspond to $\mathcal{P}=0$.}
    \label{Pnorms}
\end{figure}

Figures \ref{FAmpsFeb}-\ref{fig-gordon} show results from experiments (and numerical simulations) of the models.  In the rest of this section we discuss the experimental results.  The numerical results from the models are compared with the experimental results in Section \ref{Comparisons}.

Figures \ref{FAmpsFeb} and \ref{FAmpsMay} contain plots of the (dimensionless) amplitudes of the carrier wave and the six sidebands with largest amplitudes versus (dimensionless) $\chi$ for Expts C and D respectively.  The corresponding plots for Experiments A and B are found in Carter \& Govan~\cite{CarGov}, though they are presented in dimensional form there.  In all four experiments, the amplitude of the carrier wave decreases dramatically due to the growth of sidebands and dissipation.  By design, in Expt D, the first upper sideband, $|a_1|$, started with almost twice as much energy as the first lower sideband, $|a_{-1}|$.  In Expts C and D the magnitude of the first upper sideband decreases significantly while the magnitude of the first lower sideband increases slightly.  Note that the third lower sideband in Experiment D has a magnitude smaller than experimental error (the wave gauges measure amplitudes down to $0.005$~cm).  The curves correspond to predictions obtained from the models and are discussed in the next section.

The Benjamin-Feir index (BFI), defined by
\begin{equation} 
    \text{BFI}=\frac{2\epsilon}{\Delta\bar{\omega}/{\bar{\omega}_0}},
\label{BFI}
\end{equation}
is relative measure of nonlinearity to spectral bandwidth.  If $\text{BFI}>1$, then the wave train is unstable with respect to the Benjamin-Feir instability.  See Janssen~\cite{BFIPaper} and Serio {\emph{et al.}}~\cite{BFIPaper2} for more details on the Benjamin-Feir index.  The values of the BFI for the four experiments studied herein are included in Table \ref{DataTable}.  All four BFI values are greater than 1, so the Benjamin-Feir instability is expected to play a role in all four experiments.

Figure \ref{Mnorms} shows how (dimensionless) $\mathcal{M}$ evolves as $\chi$ increases for each experiment.  The plots show that $\mathcal{M}$ decays nearly exponentially as the waves travel down the tank, i.e.~nearly $\mathcal{M}(\chi)=\mathcal{M}(0)\exp(-2\delta\chi)$, for all four experiments.  Table \ref{DataTable} contains the least-squares best-fit values of $\delta$.  We emphasize that this empirical definition of $\delta$ combines all dissipative effects, regardless of their source, into a single term.  The mathematical models handle dissipation in a variety of ways.  In their derivation, Dias {\emph{et al.}}~\cite{DDZ} determine that $\delta=\frac{4\bar{k}_0^2\bar{\nu}}{\epsilon^2\bar{\omega}_0}$ where $\bar{\nu}=1.003*10^{-1}$~cm$^2$/sec represents the kinematic velocity of the fluid.  This result is consistent with the classical Lamb~\cite{Lamb} result for a clean surface.  vanDorn~\cite{vanDorn}, shows that the dissipation due to boundary layers along the sidewalls is given by $\delta=\frac{1}{\epsilon^2\bar{\omega}_0}\sqrt{\frac{2\bar{\nu}}{\bar{\omega}_0}}$.  The inextensible surface model of Lamb~\cite{Lamb} gives $\delta=\frac{\bar{k}_0}{\epsilon^2}\sqrt{\frac{\bar{\nu}}{2\bar{\omega}_0}}$.  Table \ref{DeltaTable} includes a comparison of these theoretical values with the experimentally measured values.  The theoretical values require the sum of the sidewall rate plus a rate due to the surface.  These sums, using the clean surface value, are also shown in Table \ref{DeltaTable}.  These predicted values are in reasonable agreement with measurements from Experiments A and C. See Henderson {\emph{et al.}}~\cite{HendersonRajanSegur} for a more detailed comparison of these and other models of dissipation in water waves.  

\begin{table}
    \begin{center}
    \begin{tabular}{|c| c c c c |}
    \hline
    $\delta$ & Expt A & Expt B & Expt C & Expt D \\
    \hline
    Experimental & $0.2640$ & $0.3223$ & $0.5021$ & $0.8129$ \\
    Clean Surface & $0.0410$ & $0.0434$ & $0.0945$ & $0.0810$ \\
    Inextensible Surface & $0.7400$ & $0.7849$ & $1.7074$ & $1.4626$ \\
    Sidewall & $0.1854$ & $0.1971$ & $0.4285$ & $0.3671$ \\
    Clean Surface \& Sidewall & $0.2264$ & $0.2405$ & $0.5230$ & $0.4481$ \\
    \hline
    \end{tabular}
    \caption{Experimentally and theoretically determined values for $\delta$, the dimensionless decay rate.}
    \label{DeltaTable}
    \end{center}
    \end{table}

Figure \ref{Pnorms} shows how (dimensionless) $\mathcal{P}$ evolves as $\chi$ increases for each experiment.  In Expt A, $\mathcal{P}$ remained more or less constant.  In Expt B, $\mathcal{P}$ decreased from a positive value to a negative value.  In Expt C, $\mathcal{P}$ for about half of its evolution, changing sign, and then increased.  In Expt D, $\mathcal{P}$ decreased for a while, then leveled off, and rose at the last gauge.  Note that Figure \ref{Pnorms} contains plots of unscaled $\mathcal{P}$.  This is different than the (exponentially scaled) plots of $\mathcal{P}\mbox{e}^{2\delta\chi}$ included in Segur {\emph{et al.}}~\cite{sh}.  So, if dissipation was the only factor, we would expect to see exponential decay of $\mathcal{P}$, but we would not expect to see $\mathcal{P}$ change sign.

Figure \ref{Omegaps} contains plots of $\omega_p$ versus $\chi$ for each experiment.  The curves are from the various models and are discussed in Section \ref{Comparisons}.  These plots show that Expt A did not exhibit FD in the spectral peak sense because $\omega_p$ was constant at all gauges.  Experiments B and D exhibited temporary FD in the spectral peak sense because $\omega_p$ decreased, but did not permanently remain lower.  However, note that it is possible that this FD would have become permanent had the experimental tank been longer.  Expt C exhibited a permanent FD in the spectral peak sense because the magnitude of the first lower sideband overtook that of the carrier wave and remained dominant.

Figure \ref{Omegams} contains plots of $\omega_m$ versus $\chi$.  The curves are from the models and are discussed in Section \ref{Comparisons}.  A comparison of Figures \ref{Omegaps} and \ref{Omegams} shows that there does not appear to be a relationship between the evolution of $\omega_p$ and $\omega_m$.  Expt A did not exhibit FD in the spectral mean sense because $\omega_m$ remained essentially constant throughout the experiment.  Expt B exhibited FD in the spectral mean sense because $\omega_m$ decreased significantly and even changed sign.  Although it is difficult to interpret the final experimental data point, Expt C exhibited FD in the spectral mean sense because $\omega_m$ decreased and changed sign.  In Expt D, $\omega_m$ started out more or less constant and then increased dramatically in the last three gauges.  This means that Expt D exhibited frequency upshift in the spectral mean sense near the end of the experiment.

\section{Comparison of model and experimental results}
\label{Comparisons}

All model equations were solved numerically using the high-order operator splitting methods introduced by Yoshida~\cite{yoshida}.  The linear parts of the PDEs were solved exactly in Fourier space using fast Fourier transforms (FFTs) and the nonlinear parts were either solved exactly in physical space (NLS, dNLS) or solved pseudospectrally using fourth-order Runge-Kutta (Dysthe, vDysthe, dGT, Gordon, IS).  Periodic boundary conditions on the (dimensionless) interval $\xi\in[0,\epsilon\bar{\omega}_0\bar{t}_f)$, where $\bar{t}_f$ is the length (in seconds) of the time series collected at the first gauge, were imposed.  The initial conditions were
\begin{equation}
B(\xi,\chi=0)=\frac{\bar{k}_0}{\epsilon}\sum_{m=-M}^{M}\bar{a}_m\exp\Big{(}im\frac{2\pi}{\epsilon\bar{\omega}_0\bar{t}_f}\xi\Big{)},
\end{equation}
where the $\bar{a}_m$ are the coefficients determined by taking an FFT of the time series recorded by the first gauge.  Only the $2N+1$ modes closest to and including the carrier wave (i.e. the modes corresponding to $1.667$ to $5.000$ Hz) were used in the initial conditions because the models are based on narrow bandwidth assumptions.  See Table \ref{DataTable} for the values of $\bar{t}_f$, $N$, and the other parameters for each experiment.

\begin{figure}
\centering
\includegraphics[width=12cm]{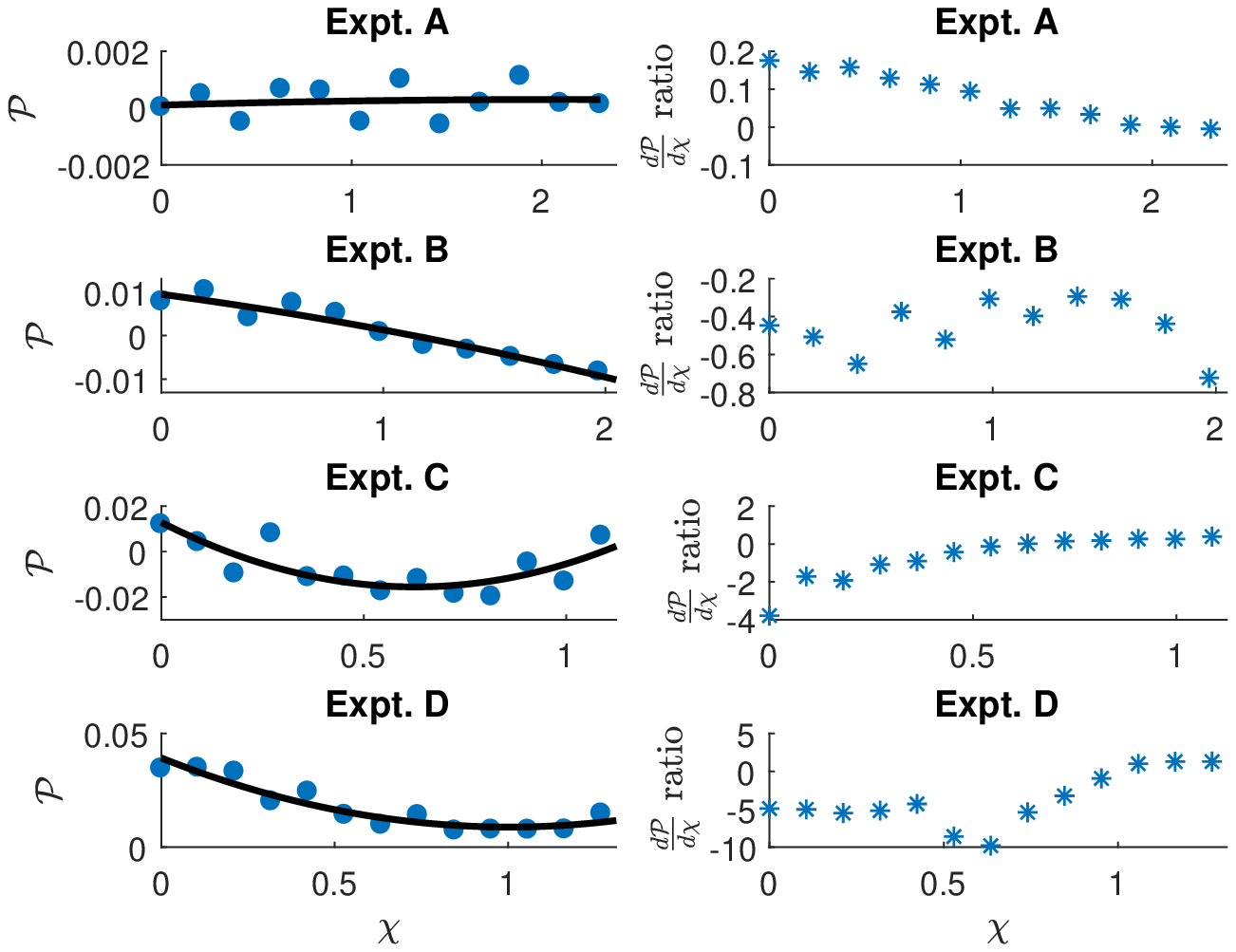}
\caption{The first column contains plots of the linear momentum, $\mathcal{P}$, versus dimensionless distance down the tank, $\chi$, for each experiment.  The dots correspond to experimental measurements and the curves are the best quadratic fits.  The second column contains plots of the ratio of experimentally determined $\mathcal{S}$, see (\ref{Sdef}), and $\frac{d\mathcal{P}}{d\chi}$ estimated using the quadratic fits shown in the first column.}
\label{fig-gordon}
\end{figure}

We start by considering the Gordon model.  First, it preserves $\mathcal{M}$ while all four experiments show that $\mathcal{M}$ decays nearly exponentially.  Second, the first column of Figure \ref{fig-gordon} contains plots of (dimensionless) $\mathcal{P}$ versus $\chi$ along with the best-fit quadratic function for each experiment.  The second column contains plots of the ratio of the experimentally computed $\mathcal{S}$ (see equations (\ref{Gordonomegamchi}) and (\ref{Sdef})) over the estimated values of $\frac{d\mathcal{P}}{d\chi}$ obtained using the quadratic functions.  Equation (\ref{GordonPchi}) establishes that if the Gordon explanation for downshifting is adequate for water waves, then this ratio would be constant in $\chi$.  Since the plots show that the ratio is not constant for any of the experiments, we conclude that one cannot find a constant $\beta_2$ so that equation (\ref{GordonPchi}) is satisfied and therefore the Gordon model for FD in optical fibers is not appropriate for FD in waves on deep water.  We note that including a term proportional to $i\delta B$ in the Gordon equation will not help, except perhaps for Expt A which has small amplitudes and does not exhibit downshifting because the Gordon term is negligible.  In Experiments B and C, the measured $\mathcal{P}$-values change sign, which cannot be explained by the model with an overall exponential decay. In Expt D, the predicted $\mathcal{P}$-value changed sign, while the measured values did not.

In order to quantitatively compare the PDE predictions with the experimental time series, we use the following (dimensionless) error norm
\begin{equation}
\mathcal{E}=\frac{1}{M-1}\sum_{m=2}^{M}\sum_{n=-12}^{12}\Big{|}\frac{\bar{k}_0}{\epsilon}\big{|}\bar{a}_n^{\text{expt}}(m)\big{|}-\big{|}a_n^{\text{PDE}}(m)\big{|}\Big{|}^2,
\label{errornorm}
\end{equation}
where $a_n^{\text{expt}}(m)$ and $a_n^{\text{PDE}}(m)$ represent the experimentally measured and numerically predicted (using the model PDE) Fourier coefficients of the $n^{th}$ mode at the $m^{th}$ gauge.  Since the time series at the first gauge is used to define the initial conditions for the PDEs, the difference between the experimental and numerical values there is zero and therefore does not need to be included in the sum.  In essence, this norm is the $L_2$ difference of the experimentally measured and numerically predicted magnitudes of the 25 Fourier modes closest to and including the carrier wave averaged across the $M-1$ downstream gauges.  These 25 Fourier modes include all of the modes shown in Figures \ref{FAmpsFeb} and \ref{FAmpsMay}.  Table \ref{ErrorTable} contains the $\mathcal{E}$ values for each of the experiments using the models discussed above.  

The Gordon and IS models contain free parameters, $\beta_1$ and $\beta_2$.  Minimizing $\mathcal{E}$ over these parameters leads to what we call the optimized IS, denoted IS$^*$, and optimized Gordon models.  The optimal Gordon model is much less accurate than dNLS, vDysthe, IS$^*$, and dGT when measuring error using $\mathcal{E}$.  Additionally, the Gordon model leads to wave breaking and poor predictions for the evolution of $\omega_p$ and $\omega_m$.  

Table \ref{ErrorTable} shows that IS$^*$ provides the minimal $\mathcal{E}$ value for Experiments B, C, and D while vDysthe provides the minimal $\mathcal{E}$ value for Experiment A.  The values of $\beta_1$ that lead to these results are included in Table \ref{BetaValues}.  As far as we can tell, there is no relation between the four $\beta_1$ values and there is no way to determine $\beta_1$ from the experimental parameters.  In contrast to the Islas-Schober assumption that $\beta_1>0$, we note that the optimal value of $\beta_1$ for Expts C and D was negative.  Figures \ref{FAmpsFeb} and \ref{FAmpsMay} shows that IS$^*$ accurately predicts the evolution of the carrier wave and the sidebands.  Figures \ref{Omegaps} and \ref{Omegams} show that IS$^*$ has a similar accuracy in predicting the evolution of the spectral peak and mean as do dNLS, vDysthe, and dGT.

Table \ref{ErrorTable} shows that the NLS and Dysthe models do not accurately model the evolution of the waves in these experiments.  Because of this, plots of the NLS and Dysthe predictions were not included in Figures \ref{FAmpsFeb} and \ref{FAmpsMay}.  They greatly over predict the growth of the first sidebands.  Additionally, NLS and Dysthe both preserve $\mathcal{M}$, which is not preserved in the experiments.

\begin{table}
    \begin{center}
    \begin{tabular}{|c| c c c c |}
    \hline   
    model & Expt A & Expt B & Expt C & Expt D \\
    \hline
    NLS & $4.05*10^{-2}$ & $8.83*10^{-2}$ & $6.05*10^{-2}$ & $8.59*10^{-3}$ \\
    dNLS & $1.19*10^{-3}$ & $3.47*10^{-2}$ & $1.02*10^{-2}$ & $5.12*10^{-3}$ \\
    Dysthe & $3.44*10^{-2}$ & $8.41*10^{-2}$ & $4.01*10^{-2}$ & $7.79*10^{-2}$ \\
    vDysthe & $6.43*10^{-4}$ & $2.33*10^{-2}$ & $8.73*10^{-3}$ & $5.39*10^{-3}$ \\
    dGT & $7.29*10^{-4}$ & $2.19*10^{-2}$ & $7.78*10^{-3}$ & $5.54*10^{-3}$ \\
    IS$^*$ & $6.75*10^{-4}$ & $6.14*10^{-3}$ & $7.72*10^{-3}$ & $4.45*10^{-3}$ \\
    \hline
    \end{tabular}
    \caption{Comparisons between experimental measurements and numerical predictions using the error norm defined in equation (\ref{errornorm}).  The errors listed for the IS$^*$ model is the minimal error found by minimizing (\ref{errornorm}) over $\beta_1$.  The optimal values of $\beta_1$ are included in Table \ref{BetaValues}.}
    \label{ErrorTable}
    \end{center}
\end{table}

\begin{table}
    \begin{center}
    \begin{tabular}{|c| c c c c |}
    \hline
    model & Expt A & Expt B & Expt C & Expt D \\
    \hline
    $\beta_1$ (IS$^*$) & $0.7$ & $32.0$ & $-2.4$ & $-20.0$ \\
    \hline
    \end{tabular}
    \caption{The $\beta_1$ values that minimize $\mathcal{E}$ for the IS$^*$ model.}
    \label{BetaValues}
    \end{center}
\end{table}

Figure \ref{Omegaps} shows comparisons of the models' predictions of the evolution of the spectral peak with the measurements. In this paragraph, every mention of FD means FD in the spectral peak sense. We note that the $n$th negative/positive scale markings on the ordinate correspond to the $n$th lower/upper sideband. 
\begin{quote}

    A. Experiment A is the small-amplitude experiment in which FD was not observed. NLS predicts a temporary FD and Dysthe predicts FD, so these models are not in agreement with experimental observations. All of the models that include dissipation agree with the experimental observations: the dNLS, vDysthe, IS, and dGT equations all predict that the spectral peak is constant. 

    B. The spectral peak in Experiment B downshifts. One data point shows a recurrence and then a return to FD. NLS predicts FD to the  first lower sideband, then the second lower sideband, followed by frequency upshift to the first upper sideband shortly after the last experimental gauge. Thus, NLS is not in agreement with observations. Dysthe predicts a temporary FD before it is observed and does not capture the, observed final FD. The dNLS, vDysthe, IS, and dGT equations all predict permanent FD and are thus in qualitative agreement with the observations. However, the dNLS, vDysthe, and dGT equations do not capture the location at which FD is observed to occur. They predict FD to occur too soon. IS* does the best job of capturing the location of the observed FD.

    C.  The spectral peak in Experiment C downshifts permanently (within the length of the tank), and FD occurs sooner than it does in Experiment B. The NLS and Dysthe equations predict the location of the FD best; however, they predict the FD to be temporary. The other models do not capture the location at which FD is observed, but do predict it to be permanent.

    D. In Experiment D the upper sideband was purposefully seeded; nevertheless, the spectral peak in that experiment downshifted. As in Experiment B, one data point shows a recurrence and then a return to FD. NLS predicts a temporary FD to the first and then to the second lower sideband, in disagreement with the observations.  Dysthe accurately predicts both the location and the permanence of the observed FD. The other models disagree with observations; they predict a constant spectral peak.
\end{quote}


Figure \ref{Omegams} shows comparisons of the models' predictions of the evolution of the spectral mean with the measurements. In this paragraph, every mention of FD means FD in the spectral mean sense. We note that both the NLS and dNLS equations preserve $\omega_m$ and therefore, predict that $\omega_m$ remains constant throughout each experiment.  The Dysthe equation predicts substantial upshifting for all four experiments.
\begin{quote}
   A. Experiment A is the small-amplitude experiment in which the spectral mean remained about constant. Therefore, the Dysthe equation necessarily does not agree with observations. The dGT equation predicts FD and so, also does not agree. The vDysthe equation predicts a slight FD near the end of the observed evolution. The NLS and dNLS necessarily agree with the observations, and here, the IS* equation also agrees.

    B. The spectral mean in Experiment B downshifts and actually changes sign.  Thus, the NLS, dNLS and Dysthe models necessarily disagree with the observations. The vDysthe model predicts a slight upshifting before FD finally takes over. Both the IS* and dGT models predict FD -- the IS* predictions are in fairly good agreement with observations, while the dGT predictions substantially over predict the measured FD. 

    C.  The spectral mean in Experiment C downshifts initially and then upshifts to just above its initial value. Thus, the NLS, dNLS and Dysthe models necessarily disagree with the observations. For this experiment, the IS* equation does not capture the initial, significant FD. Both the vDysthe and dGT equations do capture the initial FD, with the dGT predictions agreeing best, but neither of these models predict the observed ending evolution. 

    D. In Experiment D the upper sideband was purposefully seeded, and the spectral mean shows the consequences. Unlike the spectral peak, the spectral mean exhibits a permanent upshifting.  Thus, the NLS, dNLS models necessarily disagree with the observations. The Dysthe equation predicts upshifting and is in qualitative agreement. Both the vDysthe and dGT models predict FD, so they disagree with observations. The IS* equation predicts a slight upshifting, but under predicts the amount of upshifting observed.  

    For Expt D, all of the models that include linear dissipation do a reasonable job of predicting the sideband amplitudes measured; however, none was able to predict the spectral evolution as measured by either the spectral peak or spectral mean.  The only difference between Experiment D and the other experiments is that in Experiment D the upper sideband was purposefully seeded, while in the others, both the lower and upper sidebands were equally seeded.  A consequence of this forcing is that in Experiment D, all of the perturbation amplitude went into the upper sideband, while in Experiment C, which used the same value of $\bar{a}_f=r=0.5$, the perturbation amplitude was split between two modes.
\end{quote}


\section{Summary}
\label{Summary}

In summary, we have compared experimental measurements with a variety of models for the evolution of wave trains on deep water in order to gain a deeper understanding of frequency downshifting.  Table \ref{sum-results} contains a summary of our qualitative results.  We showed that the NLS equation does not accurately predict the evolution of the major Fourier amplitudes.  Nor can it predict FD in the spectral peak sense.  The Dysthe equation does not accurately predict the evolution of the major Fourier amplitudes or the spectral mean.  The Gordon equation, which comes from optics, does not accurately model the experimental data, even when optimized over its free parameter and cannot accurately model FD in water waves because of how it models the evolution of $\mathcal{P}$.  We showed that the dNLS equation accurately models the Fourier amplitudes in all four experiments, but cannot model FD in the spectral mean sense.  The viscous Dysthe equation accurately modeled the evolution of the Fourier amplitudes, reasonably predicted the evolution of the spectral peak in all four experiments, and reasonably predicted the spectral mean in three of the four experiments.  The dissipative Gramstad-Trulsen equation, introduced in equation (\ref{gvDysthe}), accurately models the evolution of the Fourier amplitudes and reasonably predicted the evolution of the spectral peak in all four experiments, and reasonably predicted the spectral mean in three of the four experiments.  Finally, the Islas-Schober equation most accurately models the experimental data in three of the four data sets when optimized over its free parameter.  Unfortunately, there is not an empirical way to determine this parameter.  All of the models are derived with either some ad-hoc term or a free-parameter, and none of the modes captures all of the features of all of the experiments, so a derivation from first principles remains an open problem.

\begin{table}
    \begin{center}
    \begin{tabular}{|c|c|c|c|c|c|c|c|}
    \hline
    Model  &  Fourier Amplitudes  &  $\mathcal{M}$ &   $\omega_p$ & $\omega_m$  \\
    ~ &  (Figures \ref{FAmpsFeb} and \ref{FAmpsMay}) &  ~ &  (Figure \ref{Omegaps}) & (Figure \ref{Omegams}) \\
    \hline
    NLS  &  0 & 0 & 0 & A \\
    Dysthe & 0 & 0 & C, D & D \\
    dNLS & A, B, C, D &  A, B, C, D  & A, B, C & A \\
    vDysthe & A, B, C, D & A, B, C, D & A, B, C & A, B, C \\
    IS* & A, B, C, D & A, B, C, D  & A, B, C & A, B, D \\
    dGT & A, B, C, D & A, B, C, D  & A, B, C & B, C \\           
    \hline
    \end{tabular}
    \caption{Qualitative summary of results.  0 means qualitative disagreement of predictions from the model (row) with the measure listed (column) for all 4 experiments. The letters correspond to experiments in which there was at least some overall qualitative agreement.}
    \label{sum-results}
    \end{center}
    \end{table}

\section{Acknowledgements}
We thank Andrea Armaroli, Bernard Deconinck, Debbie Eeltink, Paul Milewski, and Harvey Segur for helpful discussions.  This material is based upon work supported by the National Science Foundation under grants DMS-1716120 and DMS-1716159.


\begin{thebibliography}{10}

    \bibitem{AS}
    M.~J. Ablowitz and H.~Segur.
    \newblock {\em Solitons and the Inverse Scattering Transform}.
    \newblock SIAM, Philadelphia, 1981.
    
    \bibitem{Brunetti1}
    M.~Brunetti and J.~Kasparian.
    \newblock Modulational instability in wind-forced waves.
    \newblock {\em Physics Letters A}, 378:3626--3630, 2014.
    
    \bibitem{Brunetti2}
    M.~Brunetti, N.~Marchiando, N.~Berti, and J.~Kasparian.
    \newblock Nonlinear fast growth of water waves under wind forcing.
    \newblock {\em Physics Letters A}, 378:1025--1030, 2014.
    
    \bibitem{CarGov}
    J.~D. Carter and A.~Govan.
    \newblock Frequency downshift in a viscous fluid.
    \newblock {\em European Journal of Mechanics - B/Fluids}, 59:177--185, 2016.
    
    \bibitem{DDZ}
    F.~Dias, A.~I. Dyachenko, and V.~E. Zakharov.
    \newblock Theory of weakly damped free-surface flows: {A} new formulation based
      on potential flow solutions.
    \newblock {\em Physics Letters A}, 372:1297--1302, 2008.
    
    \bibitem{Dysthe}
    K.~B. Dysthe.
    \newblock Note on a modification to the nonlinear {S}chr\"odinger equation for
      application to deep water waves.
    \newblock {\em Proceedings of the Royal Society of London A}, 369:105--114,
      1979.
    
    \bibitem{Gordon}
    J.~P. Gordon.
    \newblock Theory of the soliton self-frequency shift.
    \newblock {\em Optics Letters}, 11(10):662--664, 1986.
    
    \bibitem{GT}
    O.~Gramstad and K.~Trulsen.
    \newblock Hamiltonian form of the modified nonlinear {S}chr\"odinger equation
      for gravity waves on arbitrary depth.
    \newblock {\em Journal of Fluid Mechanics}, 670:404--426, 2011.
    
    \bibitem{Hara}
    T.~Hara and C.~C. Mei.
    \newblock Frequency downshift in narrow banded surface waves under the
      influence of wind.
    \newblock {\em Journal of Fluid Mechanics}, 230:429--477, 1991.
    
    \bibitem{HendersonRajanSegur}
    D.~Henderson, G.~K. Rajan, and H.~Segur.
    \newblock Dissipation of narrow-banded surface water waves.
    \newblock {\em Fields Institute Communications}, 75:163--183, 2015.
    
    \bibitem{IslasSchober}
    A.~Islas and C.~M. Schober.
    \newblock Rogue waves and downshifting in the presence of damping.
    \newblock {\em Natural Hazards and Earth System Sciences}, 11:383--399, 2011.
    
    \bibitem{BFIPaper}
    P.~A. E.~M. Janssen.
    \newblock Nonlinear four-wave interactions and freak waves.
    \newblock {\em Journal of Physical Oceaongraphy}, 33(4):863--884, 2003.
    
    \bibitem{Kato}
    Y.~Kato and M.~Oikawa.
    \newblock Wave number downshift in modulated wavetrain through a nonlinear
      damping effect.
    \newblock {\em Journal of the Physical Society of Japan}, 64:4660--4669, 1995.
    
    \bibitem{vDystheBigger}
    O.~Kimmoun, H.~C. Hsu, B.~Kibler, and A.~Chabchoub.
    \newblock Nonconservative higher-order hydrodynamic modulation instability.
    \newblock {\em Physical Review E}, 96:022219, 2017.
    
    \bibitem{LY}
    B.~M. Lake and H.~C. Yuen.
    \newblock A note on some nonlinear water-wave experiments and the comparison of
      data with theory.
    \newblock {\em Journal of Fluid Mechanics}, 83:75--81, 1977.
    
    \bibitem{Lakeplus}
    B.~M. Lake, H.~C. Yuen, H.~Rungaldier, and W.~E. Ferguson.
    \newblock Nonlinear deep water waves: theory and experiment. {P}art 2.
      {E}volution of a continuous wave train.
    \newblock {\em Journal of Fluid Mechanics}, 83:49--74, 1977.
    
    \bibitem{Lamb}
    H.~Lamb.
    \newblock {\em Hydrodynamics}.
    \newblock Dover, New York, 1993.
    
    \bibitem{LoMei}
    E.~Lo and C.~C. Mei.
    \newblock A numerical study of water-wave modulation based on a higher-order
      nonlinear {S}chr\"odinger equation.
    \newblock {\em Journal of Fluid Mechanics}, 150:395--416, 1985.
    
    \bibitem{Ma}
    Y.~Ma, G.~Dong, M.~Perlin, X.~Ma, and G.~Wang.
    \newblock Experimental investigation on the evolution of the modulation
      instability with dissipation.
    \newblock {\em Journal of Fluid Mechanics}, 711:101--121, 2012.
    
    \bibitem{Melville}
    W.~K. Melville.
    \newblock The instability and breaking of deep-water waves.
    \newblock {\em Journal of Fluid Mechanics}, 115:165--185, 1982.
    
    \bibitem{MM}
    F.~M. Mitschke and L.~F. Mollenauer.
    \newblock Discovery of the soliton self-frequency shift.
    \newblock {\em Optics Letters}, 11:659--661, 1986.
    
    \bibitem{sh}
    H.~Segur, D.~Henderson, J.~D. Carter, J.~Hammack, C.~Li, D.~Pheiff, and
      K.~Socha.
    \newblock Stabilizing the {B}enjamin-{F}eir instability.
    \newblock {\em Journal of Fluid Mechanics}, 539:229--271, 2005.
    
    \bibitem{BFIPaper2}
    M.~Serio, M.~Onorato, A.~R. Osborne, and P.~A. E.~M. Janssen.
    \newblock On the computation of the {B}enjamin-{F}eir {I}ndex.
    \newblock {\em Nuovo Cimento-Societa Italiana di Fisica Sezione C},
      28(6):893--903, 2005.
    
    \bibitem{Suetal}
    M.~Y. Su, M.~Bergin, P.~Marler, and R.~Myrick.
    \newblock Experiments on nonlinear instabilities and evolution of steep
      gravity-wave trains.
    \newblock {\em Journal of Fluid Mechanics}, 124:45--72, 1982.
    
    \bibitem{SulemSulem}
    C.~Sulem and P.~L. Sulem.
    \newblock {\em The nonlinear {Schr\"odinger} equation}.
    \newblock Springer, New York, 1991.
    
    \bibitem{TD1990}
    K.~Trulsen and K.~B. Dysthe.
    \newblock Frequency down-shift through self modulation and breaking.
    \newblock In {\em NATO ASI Series 178}, pages 561--572, 1990.
    
    \bibitem{vanDorn}
    W.~G. vanDorn.
    \newblock Boundary dissipation of oscillatory waves.
    \newblock {\em Journal of Fluid Mechanics}, 24:769--779, 1966.
    
    \bibitem{WuLiuYue}
    G.~Wu, Y.~Liu, and D.~K. Yue.
    \newblock A note on stabilizing the {B}enjamin-{F}eir instability.
    \newblock {\em Journal of Fluid Mechanics}, 556:45--54, 2006.
    
    \bibitem{yoshida}
    H.~Yoshida.
    \newblock Construction of higher order sympletic integrators.
    \newblock {\em Physics Letters A}, 150:262--268, 1990.
    
    \bibitem{Zak1968}
    V.~E. Zakharov.
    \newblock Stability of periodic waves of finite amplitude on the surface of a
      deep fluid.
    \newblock {\em Journal of Applied Mechanics and Technical Physics},
      9(2):190--194, 1968.
    
    \end{thebibliography}
\end{document}